\documentclass[aps,eqsecnum,apsi,twocolumn,preprintnumbers,showpacs,prd]{revtex4}
\usepackage{graphicx}
\usepackage{latexsym}
\usepackage{amsmath}
\usepackage{amssymb}
\usepackage{epsfig}
\usepackage{longtable}

\newcommand{\be}{\begin{equation}}
\newcommand{\ee}{\end{equation}}
\newcommand{\bea}{\begin{eqnarray}}
\newcommand{\eea}{\end{eqnarray}}

\def\fb{\, {\rm fb}}

\def\met{E_T \hspace*{-1.1em}/\hspace*{0.5em}}
\def\lv{L \hspace*{-0.4em}/\hspace*{0.3em}}

\def\tev{\, \, {\rm TeV}}
\def\gev{\, \, {\rm GeV}}

\def\neu {{\tilde {\chi_1}}^0}
\def\cha {{\tilde {\chi_1}}^{\pm}}

\def\lslep {\tilde{e_L}}
\def\stau {\tilde{\tau_1}}

\def\mel {m_{\lslep}}
\def\mneu {m_{\neu}}
\def\mcha {m_{\cha}}
\def\mstau {m_{\stau}}

\begin{document}

\preprint {RECAPP-HRI-2010-008}

\title{\Large \bf Same-sign trileptons and four-leptons 
as signatures of new physics at the Large Hadron Collider}

\author{Biswarup Mukhopadhyaya$^{a,b}$ \footnote{E-mail:
    biswarup@hri.res.in} and Satyanarayan
  Mukhopadhyay$^{a}$ \footnote{E-mail: satya@hri.res.in}}
\affiliation{$^a$Regional Centre for Accelerator-based Particle
  Physics, Harish-Chandra Research Institute, Chhatnag Road, Jhusi,
  Allahabad - 211 019, India.\\ $^b$Indian Association for the
  Cultivation of Science, Kolkata-700 032, India. }

\begin{abstract}
  We point out that same-sign multilepton events, not given due
  attention yet for new physics search, can be extremely useful at the
  Large Hadron Collider. After showing the easy reducibility of the
  standard model backgrounds, we demonstrate the viability of
  same-sign trilepton signals for R-parity breaking supersymmetry, at
  both 7 and 14 TeV. We find that same-sign four-leptons, too, can
  have appreciable rates.  Same-sign trileptons are also expected, for
  example, in Little Higgs theories with T-parity broken by anomaly
  terms.

\end{abstract}


\maketitle

Finding physics beyond the standard electroweak theory is an important
goal of the Large Hadron Collider (LHC). However, most proposed
signals are beset with backgrounds from processes driven by the
standard model (SM) itself, and the reduction of backgrounds requires
a Herculean effort. It is by and large agreed that signals containing
leptons (electrons or muons) are helpful from this angle. Thus one
finds a lot of interest in signals comprising dileptons, trileptons as
well as final states with higher lepton multiplicity. In addition,
same-sign dileptons (SSD) are relatively background-free if the event
selection criteria are properly chosen~\cite{SSD}.

Here we stress the importance of some unexplored signals, namely,
same-sign leptons of higher multiplicity. Among these, we mainly focus
on same-sign trileptons (SS3l). In spite of the fact that the charge
of an electron or a muon can be identified with high
efficiency~\cite{TDR}, not enough attention has been paid yet to
signals with lepton multiplicity higher than two, with all of them
having the same sign of charge. Although SS3l has been discussed in
the context of top quark production at hadron colliders of
yesteryears~\cite{Barger-top}, its capacity to reveal new physics is
still a path which remains to be explored in detail.  We shall
demonstrate below that the SM backgrounds to the SS3l signal at the
LHC can be made vanishingly small. On the other hand, substantial
rates for SS3l events are predicted in some well-motivated scenarios
beyond the standard model. They are, in fact, particularly enhanced
when one has (a) lepton number (L) violation by odd units, and (b) the
presence of self-conjugate massive particles.  We illustrate this in
the context of several supersymmetric (SUSY) scenarios with R-parity
violation ~\cite{Barbier}. We also point out that other new physics
scenarios, such as Little Higgs models with T-parity broken through
anomaly terms~\cite{Hill_Hill}, can predict a signal of this
kind. Based on these observations, we conclude that framing the
experimental strategies to capture SS3l events can open a new door to
the study of physics beyond the SM at the LHC.  We further show that
such scenarios can yield enough same-sign four-lepton (SS4l) events,
which are background-free and reveal important information on the
underlying new physics.

We start by taking a look at the SM contributions to the SS3l signal.
The main sources here are (i) $t\bar{t}$, (ii) $t\bar{t}W$, (iii)
$t\bar{t}b\bar{b}$ and (iv) $t\bar{t}t\bar{t}$ production. Of the
various processes, $t\bar{t}$ production, copious as it is, generates
SS3l if a lepton comes from a charm quark produced from a $b$ which in
turn results from top-decay.  This causes a significant degradation of
momentum of at least the softest lepton, and judicious lepton
isolation and hardness cuts suppress it. The other channels, too,
suffer from either perturbative suppression at the initial production
level or low branching ratios in the cascades.  We summarise the SM
backgrounds to SS3l in Table~\ref{tab1}. The events were generated
with the code ALPGEN~\cite{ALPGEN}, and decays and hadronisation were
done using PYTHIA 6.421~\cite{PYTHIA}. We have primarily selected
leptons with $p_T \ge 10$ GeV, $|\eta| \le 2.5$, where $p_T$ and
$\eta$ are respectively the transverse momentum and pseudorapidity of
the lepton. The effect of $B^0 - \bar{B}^0$ mixing on lepton signs has
been taken into account within PYTHIA. We have approximated the
detector resolution effects by smearing the energies (transverse
momenta) of the leptons and jets with Gaussian
functions~\cite{TDR}. We further demand a lepton-lepton separation
$\Delta R_{ll} \ge 0.2$, where $(\Delta R)^2 = (\Delta \eta)^2 +
(\Delta \phi)^2$ quantifies the separation in the
pseudorapidity-azimuthal angle plane.  We also demand a lepton-jet
separation $\Delta R_{lj} \ge 0.4$ for all jets with $E_T \ge 20$ GeV.
Also, a relative isolation criterion to restrict the hadronic activity
around a lepton has been used, i.e., we demand $\sum p_T$ (hadron)
/$p_T$ (lepton)$\le 0.2$, where the sum is over all hadrons within a
cone of $\Delta R \le 0.2$ around the lepton. A missing-$E_T$ ($\met$)
cut of 30 GeV is also included, in order to reduce the probability of
jets faking leptons~\cite{Ozcan}. Subsequently, stronger $p_T$-cuts
(as mentioned in the caption of Table~\ref{tab1}) are applied, in
order to ensure minimum hardness for even the softest of the three
leptons~\cite{Sullivan}.  This, together with the demand on lepton
isolation, strongly suppresses the b(and c)-induced leptons, and makes
the SM contributions quite small, as shown in Table~\ref{tab1}.

Encouraged by the above observation, we first illustrate the
usefulness of the SS3l channel in new physics scenarios. As we have
mentioned already, our purpose is not to highlight any particular new
theory; we stress that {\em such signals, experimentally quite
  tractable as they are, speak for new physics unequivocally, and they
  are indeed expected with large rates in a number of cases}. As the
same-sign multileptons are facilitated when L-violation takes place,
R-parity violating (RPV) SUSY is our best example, where one further
has Majorana fermions in the form of gluinos and neutralinos, from
whose cascade decays leptons of either charge are expected with the
same rate. We present several cases below, with quantitative
predictions for each of them.

\begin{table}[ht]
	\centering

\begin{ruledtabular}
\begin{tabular}{c c c}

Process&${\sigma_{SS3l}}$ (fb)&${\sigma_{SS3l}}$ (fb)\\
&[Cut-1]&[$+$ Cut-2]\\
		\noalign{\smallskip}\hline\noalign{\smallskip}

 $t \bar{t} W$         &$2.80\times {10^{-2}}$&$2.44\times {10^{-3}}$ \\
\noalign{\smallskip}
 $t \bar{t} b \bar{b}$ &$4.45\times {10^{-3}}$&$<1.11\times {10^{-3}} $\\
\noalign{\smallskip}

 $t \bar{t} t \bar{t}$ &$8.40\times {10^{-4}}$&$6.45\times {10^{-5}}$\\ 
\noalign{\smallskip}

\hline 
\noalign{\smallskip}

Total  & ${3.33\times {10^{-2}}}$&${2.50\times {10^{-3}}}$\\
	\end{tabular}
\end{ruledtabular}

\caption{\label{tab1}Dominant same-sign trilepton SM background
  cross-sections ($\sigma_{SS3l}$) for $\sqrt{s}=14 \tev$ after the
  basic isolation cuts (Cut-1) and after demanding that
  $p_{T}^{l_{1}}> 30 \gev$, $p_{T}^{l_{2}}> 30 \gev$, $p_{T}^{l_{3}}>
  20 \gev$ and $E_T \hspace*{-1.1em}/\hspace*{0.5em}>30 \gev$, which
  are collectively referred to as Cut-2. Here $l_1$, $l_2$ and $l_3$
  are the three leptons ordered according to their $p_T$'s. Note that
  the $t \bar {t}$ contribution falls drastically after Cut-1 itself.}
\end{table}

{\bf {\em SS3l in RPV SUSY:}} The superpotential in RPV SUSY can
contain the following $\Delta L = 1$ terms, over and above those
present in the minimal SUSY standard model (MSSM):
 
\begin{equation}
\nonumber
W_{\lv} = \lambda_{ijk}L_iL_j{\bar{E}}_k + {\lambda^{\prime}_{ijk}}L_iQ_j{\bar D}_k+ {\epsilon}_i L_i H_2
\end{equation}

{\bf {\em Case 1}:} With the $\lambda$-type terms, we consider two
possibilities, namely, having (a) the lightest neutralino ($\neu$) and
(b) the lighter stau ($\stau$) as the lightest SUSY particle (LSP). In
(a), SS3l can arise if $\neu$ decays into a neutrino, a tau ($\tau$)
and a lepton of either of the first two families.  With the $\tau$
decaying hadronically, the two leptons from two $\neu$'s produced in
SUSY cascades are of identical sign in 50\% cases. An additional
lepton of the same sign, produced in the decays of chargino ($\cha$)
in the cascade, leads to SS3l. If there is just one $\lambda$-type
coupling (we have used $\lambda_{123}$ for illustration), there is no
further branching fraction suppression in LSP decay, and one only pays
the price of $\cha$-decay into a lepton of the same sign. In (b), two
same-sign $\stau$'s can be produced from two $\neu$'s, thanks to its
Majorana character. Each of these $\stau$'s goes into a lepton and a
neutrino; these two leptons, together with one of identical sign from
the cascade, lead to SS3l signals.

{\bf {\em Case 2}:} With $\lambda'$-type interactions, a $\neu$-LSP
decays into two quarks and one charged lepton or neutrino. If the LSP
is not much heavier than the top quark, and if the effect of the
difference between up and down couplings of the neutralino can be
neglected, we obtain SSD's from a pair of $\neu$'s roughly in 12.5\%
of the cases. If another lepton of the same sign arises from a $\cha$,
SS3l is an immediate consequence. Therefore, the overall rate of SS3l
can be sizable in this case as well. Here, (and also partially in case
1(b)), the large boost of the $\neu$ can lead to collimated jets and
leptons.  Thus some events may not pass the isolation cut.  It should
also be noted that a $\stau$-LSP with $\lambda'$-type terms cannot
lead to SS3l, as the $\stau$ decays into two quarks only.

{\bf {\em Case 3}:} With bilinear R-parity breaking terms ($\sim
\epsilon_i$), the most spectacular consequence is the mixing between
neutralinos and neutrinos as well as between charginos and charged
leptons.  Consequently, over a substantial region of the parameter
space, a $\neu$ LSP in this scenario decays into $W \mu$ or $W \tau$
in 80\% cases altogether, so long as the R-parity breaking parameters
are in conformity with maximal mixing in the $\nu_{\mu} - \nu_{\tau}$
sector~\cite{R_M_2}. From the decay of the two $\neu$'s, one can
obtain SSD's either from these $\mu$'s, or from the leptonic decay of
the $W$'s or the $\tau$'s. An additional lepton from the SUSY cascade
results in SS3l again. Adding up all the above possibilities, the
rates can become substantial.

\begin{table*}
	\centering

\begin{ruledtabular}
\begin{tabular}{c c c c c c c c c c}

Case &$\tan \beta$ & $m_{\tilde {g}}$ &$\mcha$&$\mneu$&$\mstau$& $\mel$&RPV&$\sigma_{SS3l}^1$&$\sigma_{SS3l}^2$\\
   &(GeV)&(GeV)&(GeV)&(GeV)&(GeV)&(GeV)&Coupling  &(fb) &(fb)\\
\hline
1a(1)& 15 &661  & 200 &$108^*$& 115 & 204& $\lambda_{123}$ &465.22& 195.97\\
1a(2)&40 &610 &183 & $99^*$& 139 & 265& $\lambda_{123}$& 811.20&301.36\\
1a(3)&5& 1009 &331&$176^*$&191 & 309&$\lambda_{123}$ &81.54 &55.31\\
1a(4)&40 & 1016 &337 &$178^*$&246&418&$\lambda_{123}$&55.52 &31.83\\
\hline
1b(1)& 10 & 770 & 241 & 129 & $118^*$ & 222 & $\lambda_{123}$ & 416.62 & 296.26\\ 
1b(2)& 40 & 608 & 182 & 98  & $94^*$  & 236 & $\lambda_{123}$ & 100.27 & 61.62 \\
1b(3)& 5  &1008 & 330 &176  & $171^*$ & 297 & $\lambda_{123}$ & 53.00 & 42.74  \\
1b(4)& 40 &1009 & 336 &178  & $109^*$ & 328 & $\lambda_{123}$ & 20.05 &13.41  \\
\hline
2(1)& 15 &661  & 200 &$108^*$& 115 & 204&$\lambda^{\prime}_{112}$&59.96&20.97\\
2(2) &40 &610 &183 & $99^*$& 139 & 265& $\lambda^{\prime}_{112}$ & 136.35 & 38.21\\
2(3) &5& 1009 &331&$176^*$&191 & 309& $\lambda^{\prime}_{112}$ &21.76 &12.26\\
2(4) &40 & 1016 &337 &$178^*$&246&418& $\lambda^{\prime}_{112}$& 15.27&8.21\\
\hline
3(1) &5& 1009 &331&$176^*$&191 & 309 & $\epsilon_i$ & 36.50&22.23 \\
3(2) &40 & 1016 &337 &$178^*$&246&418& $\epsilon_i$ & 23.28 &12.52\\

	\end{tabular}
\end{ruledtabular}

\caption{\label{tab2} SS3l cross-sections after Cut-1
  ($\sigma_{SS3l}^1$) and Cut-2 ($\sigma_{SS3l}^2$) at $\sqrt{s}=14
  \tev$ for the various cases discussed in the text (e.g., 1a(1)
  corresponds to the first example in case 1a). The LSP in a given
  point is indicated by a * against its mass. The low-scale MSSM
  parameters were generated in an mSUGRA framework. The $\lambda$ and
  $\lambda'$ couplings are set at 0.001, and the $\epsilon_i$ are
  within the limits set by neutrino data (see discussion in the text). }
\end{table*}

{\bf {\em Results:}} In Tables ~\ref{tab2} and ~\ref{tab3}, the
predictions for all the aforementioned cases, corresponding to some
representative points for each, are presented, for $\sqrt{s} =$ 14 and
7 TeV, respectively. We have used CTEQ6L1~\cite{CTEQ} parton
distribution functions, with the renormalisation and factorisation
scales kept at the PYTHIA default~\cite{PYTHIA}. The value of each
trilinear coupling ($\lambda, \lambda'$) used for illustration is
0.001. For case 3, The values of the $\epsilon$-parameters are chosen
consistently with the neutrino data; essentially, they are tuned to
sneutrino vacuum expectation values of the order of 100 keV, in a
basis where the bilinear terms are rotated away from the
superpotential. The values of $\epsilon_i$ are also of this order in
the absence of any additional symmetry. The exact values of
$\epsilon_i$ that correspond to points $3(1)$ and $3(2)$ in
Table~\ref{tab2} depend also on other parameters of the model, such as
the L-violating soft terms in the scalar
potential~\cite{Datta_Mukho_Roy}. However, the range of values of
these parameters is of little consequence to the neutralino decay
branching ratios. Therefore, with appropriate values of these soft
terms, $\epsilon_3\approx 100$~keV, $\epsilon_1=\epsilon_2=0$ is
consistent with all our results.

Initial and final state radiation effects as well as multiple
interactions are included in the PYTHIA simulation, where all SUSY
production processes are taken into account. We show values of SUSY
parameters at the electroweak scale (in this case it has been fixed at
$\sqrt{m_{{\tilde t}_1}m_{{\tilde t}_2}}$, where ${\tilde t}_1$ and
${\tilde t}_2$ are the two mass eigenstates of the top squarks
respectively), though they have been generated, for the sake of
economy, in a minimal supergravity (mSUGRA) scenario. Since the values
of the L-violating couplings are very small, they do not affect the
renormalisation group running of mass parameters from high to low
scale~\cite{Allanach}. We have therefore generated the spectrum using
SuSpect 2.41~\cite{SUSPECT} and interfaced it with
SDECAY~\cite{SDECAY} by using the programme SUSY-HIT~\cite{SUSY-HIT}
(for calculating the decay branching fractions of the sparticles) and
finally have interfaced the spectrum and the decay branching fractions
to PYTHIA. Also, we have neglected the role of R-violating
interactions in all stages of cascades excepting when the LSP is
decaying.
 
\begin{figure}
\begin{center}
\centerline{\epsfig{file=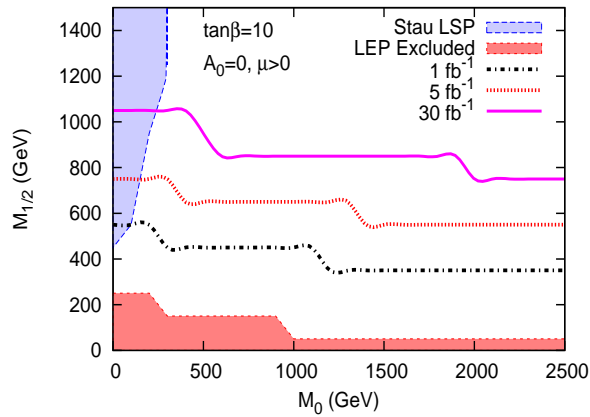,width=7.9cm,height=6.0cm,angle=-0}} 

\caption{(Color online) 10-events LHC reach with SS3l in the
  $M_0-M_{1/2}$ plane for R-parity violating mSUGRA, at $\sqrt{s}=14
  \tev$, with $\lambda_{123}=0.001$, after Cut-2.}
\label{scan}
\end{center}
\end{figure}

In Table~\ref{tab2}, we show the SS3l cross-sections for two different
gluino masses in each case, one around 600 -800 GeV, and the other in
the range of 1 TeV. We also have chosen different values of $\tan
\beta$, and made allowance for different splittings and hierarchies
between the $\cha$ and slepton masses.  For each mass range,
$\lambda_{123}$ leads to the highest rates of the SS3l signal, as in
this case the possibility of obtaining an isolated charged lepton from
the LSP decay is higher than in the two other cases. Also, if the
$\cha$'s are heavier than the first two family sleptons (and
sneutrinos), the rates go up, owing to the increase in leptonic
branching fraction of the $\cha$. Overall, the SS3l rates are
substantial for all the cases; even moderate luminosities can yield
signals for gluino masses upto a TeV or so.  In order to demonstrate
the discovery reach of the LHC in this channel, we also show in
Figure~\ref{scan}, the boundary contours of regions in the
$M_0-M_{1/2}$ plane ($M_0$ and $M_{1/2}$ being respectively the
universal scalar and gaugino mass at high scale), where at least 10
signal events can be obtained with a given integrated luminosity. This
scan was performed for a sample case (case 1) with fixed values for
the other mSUGRA parameters ($\tan \beta=10, A_0=0,\mu>0$). Similar
discovery reaches are expected for the other cases also. It should be
pointed out here that, in the scenarios we consider, the reach in the
SS3l channel is expected to be similar to the reach in channels with
higher lepton multiplicity. This is because if we assume that the
backgrounds in the multilepton channels can be reduced with similar
efficiencies as shown here for SS3l, the signal cross-sections for
four-lepton and SS3l are expected to be of similar order. While going
from trileptons to SS3l we retain 25\% of the signal, and a similar
reduction will occur while going from trileptons to four-leptons, too
(because of the $\cha\rightarrow l^{\pm} \nu \neu$ branching
fraction).

Note that there are two kinks observed in each curve of
Figure~\ref{scan}. As we increase $M_0$ for a given $M_{1/2}$, the
first two family sleptons eventually become heavier than the chargino,
thereby reducing the branching fraction of $\cha\rightarrow l^{\pm}
\nu \neu$. This leads to a drop in the SS3l cross-section, giving rise
to the first kink. The second kink is coming from the drop in the
total SUSY production cross-section as the squarks become heavier as
$M_0$ is increased, and after a certain point it is only the gluino
pair production that dominates the total cross-section. As we are
using a 10-events discovery criterion (because of negligible
backgrounds), the kinks look rather sharp.

Table~\ref{tab3} shows the points where we can get at least 10 signal
events even at 7 TeV within an integrated luminosity of $2 \fb^{-1}$.
The total SM background here, after both the cuts listed in
Table~\ref{tab1}, is $7.01 \times 10^{-4} \fb$. While the $\lambda'$
couplings lead to moderate rates here, rather handsome rates are
predicted with $\lambda$-type ones, with both the $\neu$ and $\stau$
as the LSP.  Thus we conclude that the prospect of discovering new
physics in the SS3l channel in a background-free manner is rather
bright even during the early run of the LHC.

\begin{table}[ht]
 \centering
\begin{ruledtabular}
 \begin{tabular}{c c c}
Case  & $\sigma_{SS3l}^1$&$\sigma_{SS3l}^2$\\
    & (fb)             &(fb)             \\
\hline
1a(1) & 52.64 &19.82 \\
1a(2) & 90.27 &29.45 \\
1b(1) & 44.30 &30.74 \\
1b(2) & 9.92  &6.46  \\
 \end{tabular}

\end{ruledtabular}
\caption{\label{tab3} SS3l cross-sections after Cut-1
  ($\sigma_{SS3l}^1$) and after Cut-2 ($\sigma_{SS3l}^2$) at
  $\sqrt{s}=7 \tev$ for cases as defined in Table~\ref{tab2}. The RPV
  coupling in all the above cases is $\lambda_{123}=0.001$.}
\end{table}

{\bf {\em Same-sign four-lepton (SS4l) signal:}} In all the cases
discussed above, owing to the Majorana nature of the gluino, it is
possible to produce two $\cha$'s of the same sign in an event. Thus,
in addition to SS3l, one can also have four leptons with identical
charge, coming from these two $\cha$'s and two LSP's.  Such an SS4l
signal has negligible backgrounds within the SM, particularly when
strong isolation and lepton $p_T$ cuts are used to suppress the rate
of leptons coming from heavy flavour decays. Though a further
branching fraction suppression will reduce this signal as compared to
SS3l, we note in Table~\ref{tab4} (in case 1 for illustration) that
the event rates can still be quite sizable at the LHC, during the 14
TeV run, within an integrated luminosity of $5 \fb^{-1}$.

\begin{table}[ht!]
 \centering
\begin{ruledtabular}
 \begin{tabular}{c c c}
Case  & $\sigma_{SS4l}^1$&$\sigma_{SS4l}^2$\\
    & (fb)             &(fb)             \\
\hline
1a(1)   & 15.74  &4.52  \\
1a(2)   & 33.23  &9.97  \\
1a(3)   & 4.75   &2.70  \\
1a(4)   & 3.31   &1.49 \\
1b(1)   & 24.70  &15.11 \\
1b(3)   & 2.77   &2.08  \\
 \end{tabular}

\end{ruledtabular}
\caption{\label{tab4} SS4l cross-sections after Cut-1
  ($\sigma_{SS4l}^1$) and after Cut-2 ($\sigma_{SS4l}^2$) at
  $\sqrt{s}=14 \tev$ for cases as defined in Table~\ref{tab2}. For
  SS4l Cut-2 refers to demanding lepton $p_T>20 \gev$ for all the four
  leptons and a $\met>30 \gev$. The RPV coupling in all the above
  cases is $\lambda_{123}=0.001$.}
\end{table}

{\bf {\em SS3l in Little Higgs:}} Finally, we would like to point out
that the SS3l signal is also possible in other scenarios of new
physics. An example is the Littlest Higgs model~\cite{Littlest_Higgs}
with T-parity (LHT) violated via the Wess-Zumino-Witten anomaly
term~\cite{Hill_Hill}. In this case, the heavy photon ($A_H$) (which
in most models is the lightest T-odd particle) may decay into a
$W^+W^-$ pair.  Pair-produced heavy quarks ($q_H$) can thus lead to
four W's, two of which can decay leptonically to give same sign lepton
pairs. The third additional lepton can easily come from the cascade
via the decay of the heavy partner of the W boson ($W_H$). Thus we
find that in the region of LHT parameter space where $M_{A_H}>2M_W$
and $M_{q_H}>M_{W_H}$ one can have a SS3l signal. This, in fact, is a
large region in the two-dimensional ($f,\kappa_q$) parameter space
determining the heavy quark and gauge boson masses in LHT. In
addition, if the T-odd leptons ($l_H$) are lighter than $W_H$, the
SS3l rates will be further enhanced. This is achievable within this
framework for appropriate values of $\kappa_l$. As an example, we have
generated events for the parameter choices $f=1150 \gev$,
$\kappa_q=0.5$ and $\kappa_l=0.25$, which correspond to $M_{q_H}= 809
\gev$, $M_{A_H}= 174 \gev$, $M_{W_H}=747\gev$ and $M_{l_H}= 407 \gev$
(the subscript H denotes T-odd partners of SM particles), with CalcHEP
2.5~\cite{CalcHEP,LHT} and interfaced them with PYTHIA.  We obtain an
SS3l cross-section of 3.34 fb at $\sqrt{s}=14 \tev$, after Cut-2 as
defined before.

In conclusion, same-sign multilepton signals are quite striking from
the angle of new physics search at the LHC, including its $7 \tev$
phase. Such signals can have large rates {\em if more than one
  self-conjugate particles occur in a new physics scenario. This
  feature is better reflected in SS3l and SS4l than in SSD or general
  four-lepton signals}. We have shown that clearly discernible rates
for same-sign trileptons are expected over large regions of the
parameter space of R-parity violating SUSY with broken L, even with
moderate integrated luminosity. SS4l events, too, can have substantial
rates in such scenarios. We also note that similar signals arise in
other new physics proposals, such as Little Higgs theories with
T-parity broken by anomaly terms. Due attention to this class of
signals at the LHC is therefore a desideratum.

SM thanks Nishita Desai for useful comments on the manuscript and the
Indian Association for the Cultivation of Science, Kolkata for
hospitality. This work was partially supported by funding from the
Department of Atomic Energy, Government of India for the Regional
Centre for Accelerator-based Particle Physics, Harish-Chandra Research
Institute (HRI). We also acknowledge the cluster computing facility at
HRI (http:/$\!$/cluster.hri.res.in).

\end{document}